\documentclass{article} 
\usepackage{bris}
\usepackage{psfig}  \usepackage{epsfig}
\newcommand{\lbabar}{\mbox{{\large B}{\normalsize\hspace{-0.45em} A}\hspace{-0.1em}{\large B}{\normalsize \hspace{-0.45em} A\hspace{-0.1em}R}}}
\newcommand{\babar}{$\mbox{\sl B\hspace{-0.4em} {\scriptsize\sl A}\hspace{-0.4em} B\hspace{-0.4em} {\scriptsize\sl A\hspace{-0.1em}R}}$}
\newcommand{\mum} {\ensuremath{\,\mu\mathrm{m}}}

\newcommand{\ns}  {\ensuremath{\,\mathrm{ns}}}
\newcommand{\Hz}  {\ensuremath{\,\mathrm{Hz}}}
\begin{document}
\begin{titlepage}{BRIS/HEP/2000--03}{July 2000}
\title{The \lbabar\ Calorimeter Light Pulser System} 
\author{P.J. Clark\Instref{Bristol}}
\Instfoot{Bristol}{On behalf of the \babar\ collaboration.}

\begin{abstract}
  To make precision measurements with a CsI(Tl) calorimeter in a high
  luminosity environment requires that the crystals are well calibrated
  and continually monitored for radiation damage. This should not effect
  the total integrated luminosity which is particularly important for
  the \babar\ calorimeter to enable it to make $C\!P$ violation
  measurements in the $B$ meson system. To achieve this goal a
  fibre-optic light pulser system was designed using xenon flash lamps
  as the light source.  A novel light distribution method was
  developed using an array of graded index microlenses. Some initial
  results from performance studies are presented.
\end{abstract}
\end{titlepage}
\section{Introduction}
\label{sec:introduction}
The light pulser system monitors short term changes in the response of
the \babar\ calorimeter~\cite{tdr}. It also provides a useful and flexible
diagnostic tool for the entire readout chain of the calorimeter from
light collection through to data acquisition.  The system is designed to
have a stability of 0.5\% over a period of around one week. It can also
monitor long term changes in the calorimeter response to an accuracy of
1\% using a reference system which cross-calibrates the intensity of the
light pulses against a radioactive source.
\begin{figure}[htbp]
  \begin{center}
    \includegraphics[height=100mm]{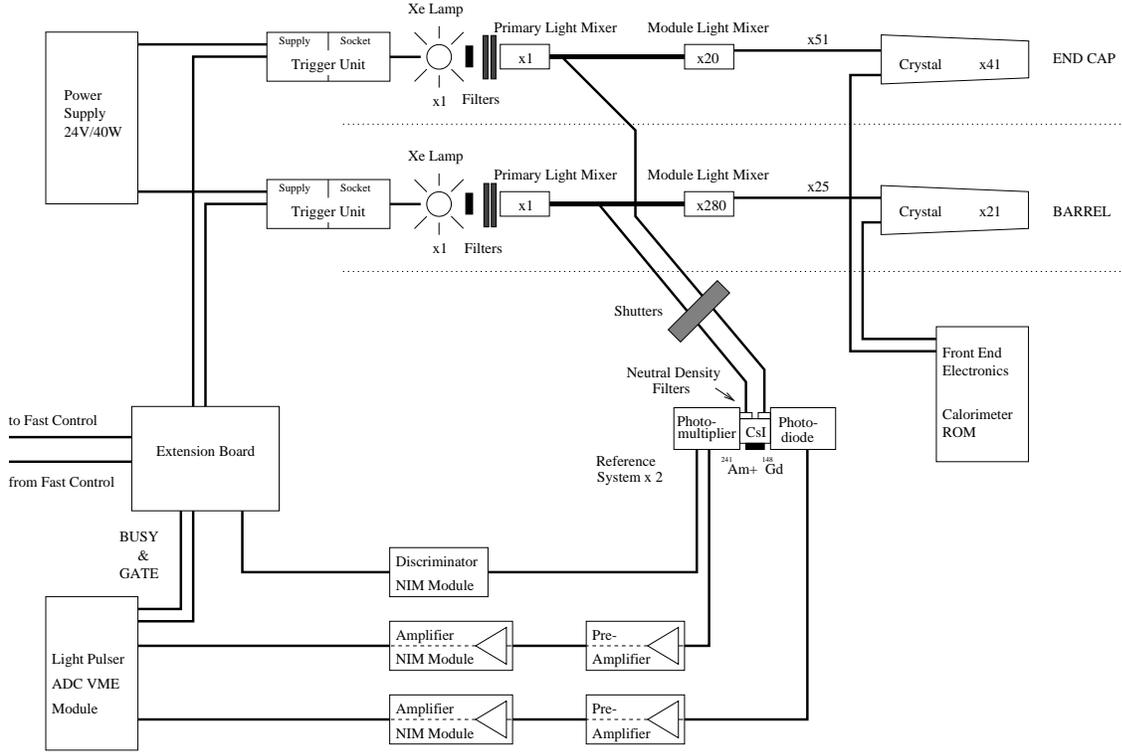} 
    \caption{Overview of the light pulser system}
    \label{fig:lps}
  \end{center}
\end{figure}

The overall design of the light pulser system is shown in
figure~\ref{fig:lps}. There are two high stability xenon flash lamps
manufactured by Hamamatsu (type L4633). One of the lamps supplies the
barrel and the other supplies the endcap calorimeter.  The light
produced from the xenon lamps is spectrally filtered to match the
emission spectrum of CsI scintillation light.  It is then attenuated by
two neutral density filters to allow the correct equivalent energy in
the calorimeter crystals to be selected.  The light fills a light mixer
bar which uniformly illuminates a bundle of 400\mum\ multi-mode fibres
which deliver the light to the individual calorimeter modules. At each
module there is a module mixer which takes the light from each fibre and
illuminates a close-packed bundle of 200\mum\ fibres which transport
light to individual CsI crystals. The light is injected into the rear
face of each crystal and then diffusely reflects within the crystal.
This effectively imitates the crystal scintillation light produced by
the energy deposition of an electromagnetic shower.  The light is then
readout using the full \babar\ calorimeter electronic readout chain from
the photodiodes right through to the data acquistion.  The light pulser
system produces an equivalent energy in the calorimeter which is high
enough to allow it to be run with beam backgrounds in the detector. A
more detailed description is given elsewhere~\cite{clark}.

There is a reference system to take out instabilities in the light
source.  Fibres are routed from the primary mixer systems to the
reference system.  The reference system itself is cross-calibrated
against a mixed alpha source ($^{148}\mathrm{Gd}$ and
$^{241}\mathrm{Am}$).  Both the alpha source and the reference fibres
are attached to a small CsI crystal which is readout by a
photomultiplier tube and a photodiode.  The reference system data from
both source events and light pulses is collected in the multi-event
buffer of a 12 bit peak sensing ADC (CAEN V556S). The source events are
accumulated continuously in the buffer and readout using a 1\Hz\ 
software trigger. The readout of light pulser system events is triggered
from a signal which is derived from the photomultiplier tube in the
reference system.  This is more accurate than using the input signal to
the lamp trigger power supply because there is a time jitter of 200\ns\ 
in the lamp response with respect to the trigger.  The lamps are
triggered at a frequency of 14.2\Hz\ which comes from dividing the
\babar\ clock.
\section{Graded index microlenses}
\begin{figure}[htbp]
  \begin{center}
    \includegraphics[width=100mm]{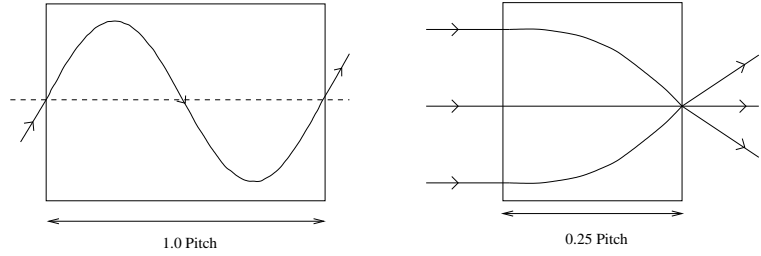}
    \caption{The principle of the microlens}
    \label{fig:microlens}
  \end{center}
\end{figure}
In the endcap system graded index microlenses (Newport LGI630-3) are
used to ensure that all the light modes in the 400\mum\ fibres are filled
correctly. This led to a better uniformity in the energy distribution.
Conventional fibre optics rely upon having a step refractive index where
the fibre core is at a higher refractive index than the cladding.
Graded index fibre optics however do not have a separate core and
cladding. Instead they have a refractive index that varies radially.
This results in an optical ray following a sinusoidal path down the
fibre rather than discrete total internal reflections, see
figure~\ref{fig:microlens}.  An interesting feature of these fibres is
if they are cut at specific lengths they act as minature lenses. If an
optical ray completes one sinusoidal oscillation within the lens it is
said to have a pitch of 1.0.  For our case lenses with a pitch of 0.25
allow a collimated light source to be focused. This is ideal for input
into a fibre optic. Each lens is coated in an anti-reflection coating
(MgF$_{\mathrm{2}}$) to maximise its transmission. The type of lens
which was chosen was plano-plano meaning that the lens is terminated
perpendicularly at both ends.

\section{Stability of the system}
\begin{figure}[htbp]
  \begin{center}
    \includegraphics[height=90mm]{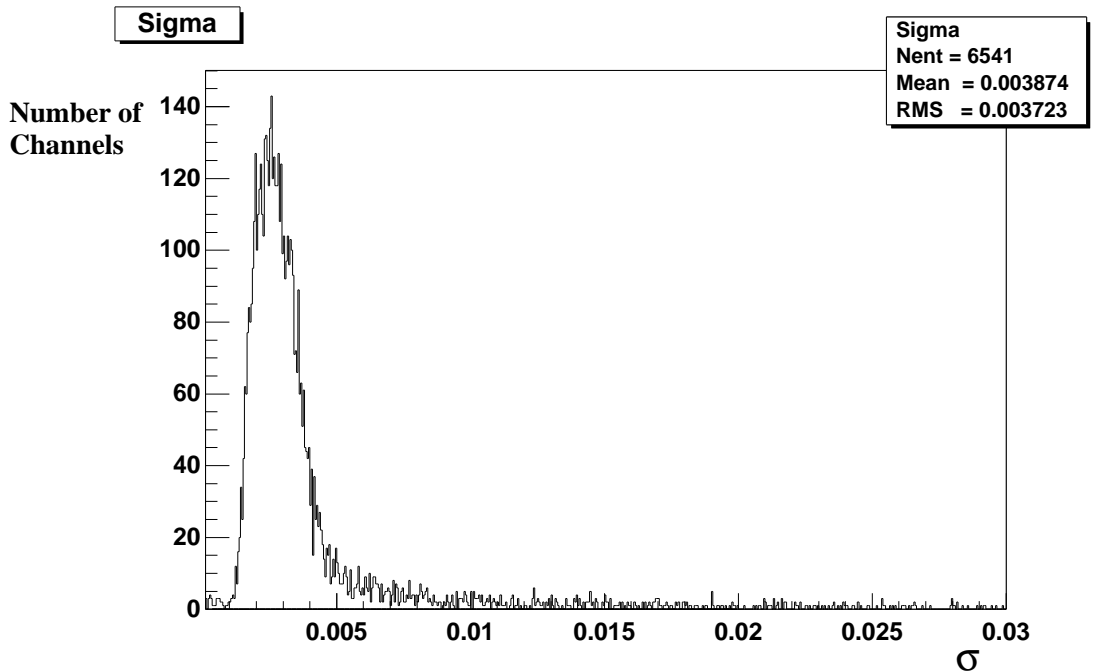}
    \caption{The stability of the light pulser system}
    \label{fig:sigma}
  \end{center}
\end{figure}
It is important for the light pulser system to have as high a
pulse-to-pulse stability as possible. In figure~\ref{fig:sigma} the
stability of the system is shown. This is produced by a Gauss fit to the
energy distribution in each channel. The data is normalised to the mean
energy in each calorimeter module on a pulse-to-pulse basis. There are a
small number of noisy channels due to preliminary electronics problems.

\section{Summary}

\begin{figure}[htbp]
  \begin{center}
    \includegraphics[height=130mm]{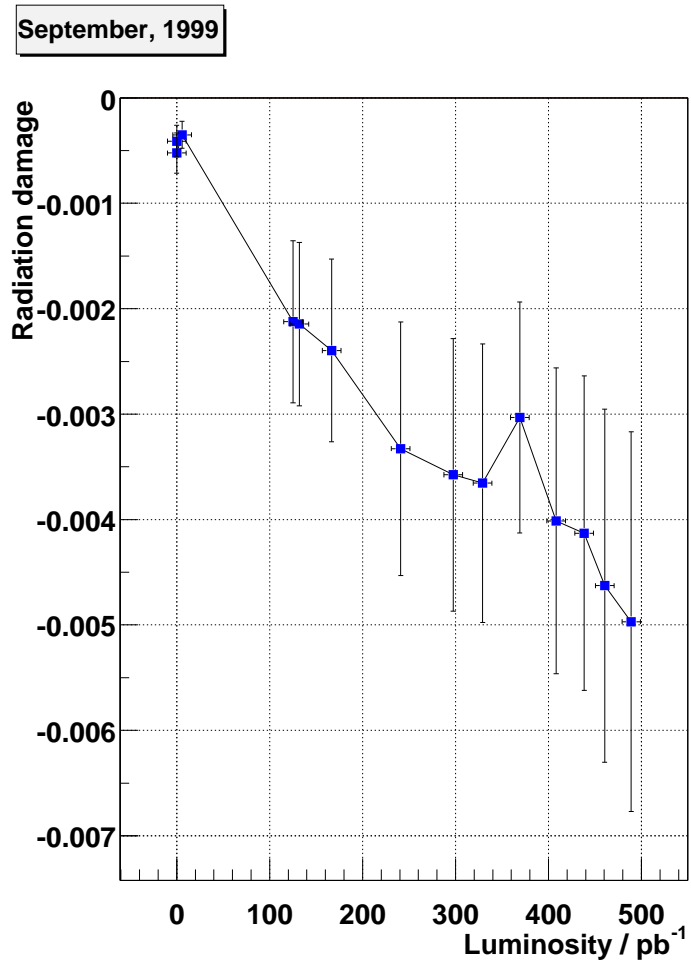}
    \caption{Differential radiation damage in the endcap versus total
      integrated luminosity}
    \label{fig:rad_damage}
  \end{center}
\end{figure}

Using the light pulser data from the inner ring of crystals in the
endcap calorimeter, where maximum radiation damage occurs, it is
possible to obtain a correlation between the total integrated luminosity
of data taken by \babar\ and the differential radiation damage in the
endcap. This is the radiation damage to the inner ring with respect to
the rest of the endcap.  The correlation is shown in
figure~\ref{fig:rad_damage}. This demonstrates that the 
light pulser system can measure accurately very small changes in the
response of the \babar\ calorimeter.
\def\bibname{References}
\bibliographystyle{plain}
\bibliography{elba}

\end{document}